# The Internet of Things in Ports: Six Key Security and Governance Challenges for the UK

**Policy Briefing**
**May 2019**

## Background

In January 2019, the UK Government published its *Maritime 2050: Navigating the Future* strategy. In the strategy, the government highlighted the importance of digitalisation (with well-designed regulatory support) to achieving its goal of ensuring that the UK plays a global leadership role in the maritime sector. Ports, the gateways for 95% of UK trade movements, were identified as key sites for investment in technological innovation. The government identified the potential of the Internet of Things (IoT), in conjunction with other information-sharing technologies, such as shared data platforms, and Artificial Intelligence applications (AI), to synchronise processes within the port ecosystem leading to improved efficiency, safety and environmental benefits, including improved air quality and lower greenhouse gas emissions.

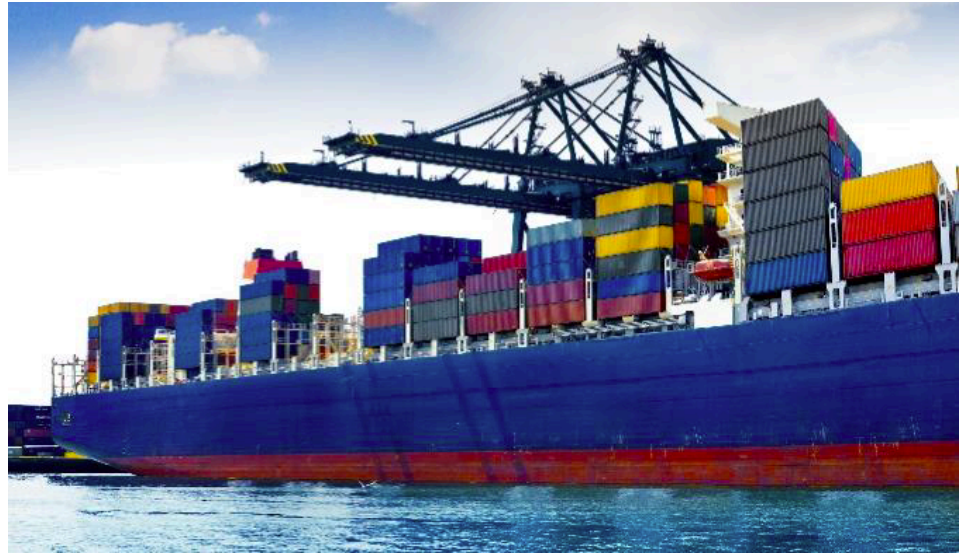

Source: *Maritime 2050, Navigating the Future (UK DfT)*

> "IoT is not "a thing", rather, it is many different "things" connected by a network that involves a vast array of application areas that extend well beyond critical infrastructures such as ports."
>
> Source: L. Tanczer et al (2018)

To achieve the desired objectives the government, businesses and other port users will need to understand the security risks and governance challenges that ensue from incorporating digital technologies at scale within the ports - the central nervous system of maritime logistics infrastructure.

**About the Authors:**

**Dr. Feja Lesniewska** is a research fellow on the EPSRC-funded PETRAS National and International Policy for Critical Infrastructure Cybersecurity (NIPC) project.

**Dr. Uchenna D Ani** is a research fellow on the EPSRC-funded PETRAS Analytical Lenses for Internet of Things Threats (ALIoTT) project.

**Professor Jeremy M Watson** is the Director of PETRAS IoT Research Hub, and Principal Investigator in the ALIoTT project

**Dr. Madeline Carr** is the Director of the Research institute in Science of Cyber Security (RISCS), and Principal Investigator in the NIPC project.

**About this Briefing:**

This briefing is based on work funded by EPSRC and carried out by UCL STEaPP NIPC-ALIoTT collaborative projects under the PETRAS Cybersecurity of the IoT Hub.

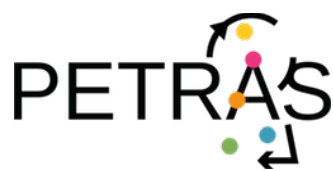

**DEPARTMENT OF SCIENCE, TECHNOLOGY, ENGINEERING AND PUBLIC POLICY**

## Our Research

This briefing is based on a Roundtable held at Gard UK Ltd, London on 8th March 2019. Participants, including representatives from academia, business, government and the insurance sector, came together to identify the challenges to creating a safe, secure, and sustainable digital port environment in the UK in line with the *Maritime 2050: Navigating the Future* strategic vision.

## The following six key security governance challenges were identified in the Roundtable

### 1. Adaptive Risk Management

An increase in the type and number of digital technologies utilised in port operational environments makes oversight difficult. There is no single security risk management approach that can be applied with confidence. The technology and associated risk dynamics necessitates an administrative procedural review that is adaptive, transparent, reliable and sustainable (Boyes et al, 2016).

The international shipping and port security (ISPS) code is an administrative security risk management tool. This tool has been developed in such a manner that it can be amended to accommodate evolving threats and risks (ISPS, 2002).

The ISPS port security standards form the basis for the Institution of Engineering and Technology (IET) / Department for Transport (DfT) port cybersecurity guidelines. The IET/DfT guidelines offer a non-legally binding set of principles for port authorities and users to meet the ISPS standards in regard to cyber technologies. (Boyes et al, 2016). The key question that remains is whether the ISPS standards, even with supporting guidelines such as those of the IET, offer a level of systems security in a rapidly changing technological port environment that will meet the needs of users, operators and insurers?

### 2. Interoperability of IoT and Legacy Systems

Interoperability among components and systems help to mitigate cybersecurity risks by closing up susceptibilities that may be enabled by seamless component interactions and exchanges. A major challenge to delivering a comprehensive risk management model for a port is the lack of interoperable information technology systems. A port needs to maintain several key attributes. These include:

1. Speed and efficiency of port operations;
2. Ability for safe port operations;
3. Health and safety of port staff and other users; and
4. Integrity of the port's physical environment.

To maintain and deliver the key attributes, many ports in the UK still rely on traditional operational technologies, such as early industrial control systems that can be vulnerable to cyber threats. The IoT will further add new layers of complexity to a legacy of established systems already installed and in operation at a port. The interfaces between older and newer digital systems require re-organisation for port authorities, and other stakeholders including terminal operators, ship operators, IT providers, borders and custom agencies. These interfaces can create new risks, many yet to be perceived and/or predicted. The capacity of a port to address novel security risks depends significantly on knowledge, capacity, and resources.

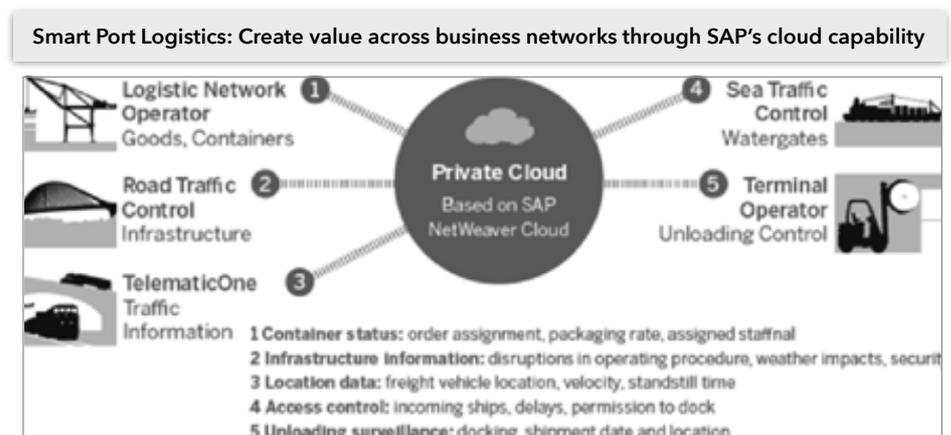

Smart Port Logistics: Create value across business networks through SAP's cloud capability

Source: *Mladen Jardas et al, (2018)*

## 3. Security Management Accountability

Under the ISPS Code, port security responsibilities are clearly outlined. It is crucial to ensure that those responsible for protecting the port facility, vessels within ports (when docked or berthed), persons, cargo, cargo transport units and ship's stores from the risks of security incidents can understand, perceive, and respond appropriately to mitigate risks. Most ports have established port security committees to coordinate implementation with the Maritime Security Measures.

Incorporating IoT and other digital technology systems into port environments makes the task of identifying those responsible for various aspects of information security and safety pressing. The person or organisation accountable for security measures is required to maintain a system that can extend beyond the physical boundaries of the port, such as data storage facilities. This is because the implications of IoT connected systems often go beyond previously understood boundaries which can challenge the UK National Cyber Security Strategy 2016-2021, (Cabinet Office, 2016) and the environment strategy (DEFRA, 2018), and their regular reviews, will increase the likelihood of the success of the UK *2050 Maritime Strategy*. Improving alignment with these and other existing and emerging strategies will enable knowledge transfer and foster targeted skills development in critical sectors.

## 4. Due Diligence and Capacity

The knowledge and skills required to maintain traditional operational technologies, such as early industrial control systems differ from those with real time data exchange through IoT, AI and 5G connections.

Once accountability is identified there is a need to understand the capacity of all port users to deliver, as outlined within the procedures, the due diligence each actor has to shoulder. As responsibility for systems moves beyond the physical boundaries of the ports, recognising who is accountable for upholding the maintenance of systems is crucial.

At the moment, responsibility is outlined in legislation and the ISPS Code. However, each attribution of responsibility needs to be reviewed on a rolling basis as the process of identifying accountable parties will be fluid as the system becomes more complex.

The transfer of data and analysis to centralised systems can obscure attribution for problems with integrity and security. Shared platforms such as the Port Community System's Marine Traffic web platforms and Electronic Data Interchange reports need to be continually maintained to avoid unauthorised access to, or destruction of, critical information.

Lessons learnt from addressing security related matters in one area need to be made visible and, where appropriate, adopted in other sectors. Through its cyber strategy, the UK government has developed and promoted important regulatory initiatives, such as the voluntary Secure by Design Code of Practice for consumer devices. There needs to be a clear approach that brings together lessons learnt rapidly across all sectors if vulnerabilities in security are to be addressed.

The *Maritime 2050 Strategy* is the first ever maritime review by the UK Government. Given the importance of the sector to the UK economy, environment, and security, a regular review, as already in place for defence, would be sensible. It would improve understanding in government, as well as in business and for the general public, of the importance of the sector. These include ports and how best to invest strategically to build a smart, sustainable future.

## 5. Alignment and

Knowing the state of play is important for any successful implementation process of a strategy.

Alignment with complimentary government strategies such as the UK Transport Strategy (DfT, 2018),

**For further information please contact:**

Dr. Feja Lesniewska
f.lesniewska@ucl.ac.uk

Dr. Uchenna D Ani
u.ani@ucl.ac.uk

## 6. The Business Case

In the UK, the business case for investing into new technologies is challenging due to variations in both the types of ports and existing ownership models.

There are a range of port business models in the UK that provide different functions and services. These include *"ro-ro"* (roll on, roll off) and passenger ferry services, containerised traffic, dry bulk cargoes (such as aggregates), liquid bulk (including oil and liquefied natural gas), the cruise industry, fishing fleets and general cargo. Some ports are also bases for vessels constructing or servicing offshore energy facilities. (Boyes, 2016).

In the UK, there are a range of port ownership models. The majority of the country's four hundred ports operate on a commercial basis without public support, in competition with rival ports (both domestically and abroad). Given the variation in political and economic contexts, each port will assess the costs and benefits of investing in IoT systems individually.

The different port models have implications for the development of a business case for the adoption and secure integration of the IoT into the operating environment. The opportunities available to a port's users depend on the capacity to adopt new technologies but also the economic opportunities that doing so will actually create.

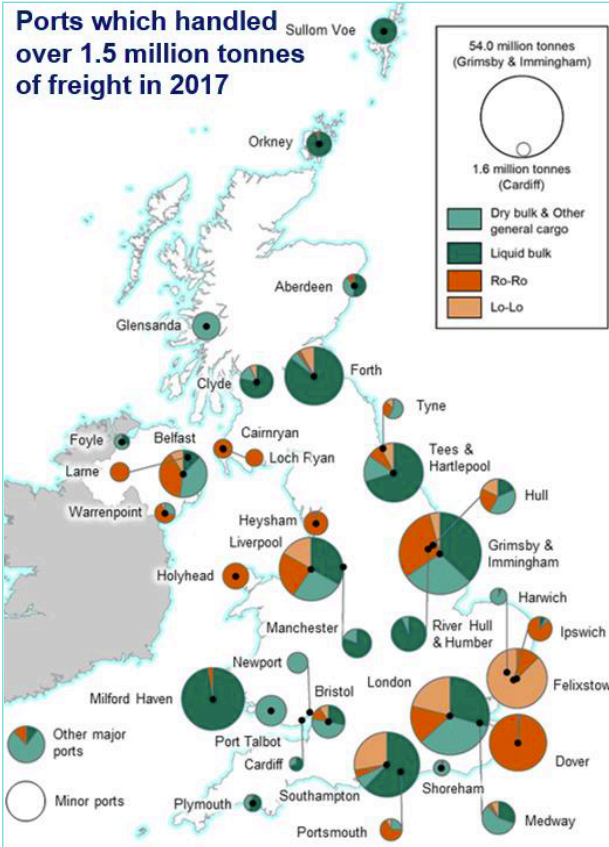

Source: Port freight statistics, DfT. © Crown Copyright. All rights reserved DfT 2018.

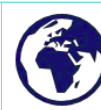 **95%** of UK imports and exports are moved by sea (approx.)

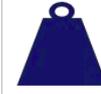 **482 million tonnes** of international and domestic freight passed through UK ports in 2017

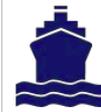 **116,000 cargo ships** arrived at UK ports in 2017

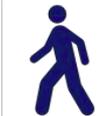 **65 million passengers** (int'l and domestic) passed through UK ports in 2017

## Conclusion

For the digital transformation in the UK ports landscape to be successful at delivering the economic, social and environmental benefits, each of the six challenges outlined in this briefing will need to be addressed collectively. As a first step, systematically reviewing the *ISPS Code* to determine where the gaps are and which changes may be needed to establish clear guidance to support those responsible for providing security within digitalised port environments will help. This will provide a foundation for developing appropriate bespoke policy and regulatory tools for digitalised port environments that will minimise security threats and deliver sustainable governance outcomes.

### References.

### Acknowledgements

This brief is part of the EPSRC PETRAS Programme National and International Critical Infrastructure Protection research project. We thank all the research Roundtable participants for their time and contributions that made this briefing possible. We are greater to Gard UK Ltd for their generosity in hosting the Roundtable.